# *Ab-initio* study of structural, elastic, electronic, optical and thermodynamic properties of MgV$_2$O$_6$

Md. Atikur Rahman[1], Md. Zahidur Rahaman[2], Md. Abdur Razzaque Sarker*

[1,2] Department of Physics, Pabna University of Science and Technology, Pabna-6600, Bangladesh
* Department of Physics, Rajshahi University, Rajshahi, Bangladesh

## Abstract

We have performed *ab*-initio calculations using plane-wave ultraviolet pseudopotential technique based on the density-functional theory (DFT) to study the structural, mechanical, electronic, optical and thermodynamic properties of orthorhombic MgV$_2$O$_6$. The calculated lattice parameters are in good agreement with the available experimental data. The second-order elastic constants and the other relevant quantities such as the Young's modulus, shear modulus, Poisson's ratio, compressibility, anisotropy factor, sound velocity, and Debye temperature have been calculated. After analyzing the calculated elastic constants, it is shown that the compound under study is mechanically stable. The analysis of the electronic band structure shows that this compound reveals semiconducting nature with band gap 2.195 eV and the contribution predominantly comes from O-2s states. Furthermore, in order to clarify the mechanism of optical transitions of MgV$_2$O$_6$, the complex dielectric function, refractive index, reflectivity, absorption coefficient, loss function and the photoconductivity are calculated and discussed in details. The large reflectivity of the predicted compound in the low energy might be useful in good candidate material to avoid solar heating. Also we have calculated the Debye temperature $\theta_D$, using our elastic constants data.

**Keywords**: *Ab*-initio calculations; elastic properties; electronic structure; optical properties; Debye temperature.

---

*Corresponding author: sarker_riphy@yahoo.com

1. Introduction

The family of vanadium oxide materials has attracted a huge attention in the research community recently due to their interesting ionic, electronic and physical characteristics. These compounds have many potential applications. These compounds can be used as cathode which is used in battery. These compounds are also suitable for electrochromic devices [1-5]. In the family of vanadium oxide the Co-based vanadium oxide compounds are the subject of great interest in condensed matter physics and material science due to their many attractive characteristics including strong anisotropic character [6-9], magnetic field induced transition [6, 10], quantum criticality nature [11-13] etc. Due to these remarkable characteristics not only Cobalt based but also Ni, Mg, Zn, Fe, W and Te-based vanadium oxides are synthesized and characterized in different time. Though a considerable amount of progress has been done to study the different physical and chemical nature of these compounds, however there is still lack of knowledge about the complete behavior of these oxides. The general structure of vanadium oxide family is $AV_2O_6$ (where, A = Co, Mg, Ni, Zn, Fe, W and Te). It possesses the $FeTa_2O_6$ type (mineral tapiolite) crystal structure. In 1997 N. Kumada *et al.* first reported the trirutile type bismuth oxide [14].

However, magnesium is a very crucial engineering material due to their huge application in practical life. It possesses low density, good stiffness and high mechanical strength. Due to these characters Mg-based alloys are used in aerospace manufacturing and automotive industry [15]. Abundance of magnesium in earth is another reason of attraction on the Mg-based alloys. Hence it is very interesting to study the Mg-based vanadium oxide. The first structural determination of $MgV_2O_6$ was reported by Gondrand M *et al.* in 1974 [16]. In 2013 Lian *et al.* synthesized $MgV_2O_6$ using HPHT (High Pressure and High Temperature) synthesis method [17]. They studied the detailed structural properties of $MgV_2O_6$ under high pressure using Raman spectroscopy at room temperature [17]. They predict a columbite-type layered structure of $MgV_2O_6$.

Though experimental characterization of the structural properties of $MgV_2O_6$ has been done there is no detailed theoretical characterization of $MgV_2O_6$ in literature. So in order to improve the theoretical insight about this compound in this present literature we aim to perform a comprehensive theoretical investigation on this compound. By using the Density

Functional Theory based CASTEP computer code we study the structural, mechanical, electronic, optical and thermodynamic properties of orthorhombic MgV$_2$O$_6$. Finally a reasonable comparison of our evaluated data with the experimental results has been made with thorough discussion.

## 2. Computational methodology

All the calculations were performed with CASTEP code [18] based on the framework of density functional theory (DFT) using a plane-wave pseudopotential method. The electronic exchange-correlation is treated under the generalized gradient approximation (GGA) in the method of Perdew–Burke–Ernzerhof (PBE) [19]. The basis set of valence electronic states was set to 2p$^6$ 3s$^2$ for Mg, 3s$^2$ 3p$^6$ 3d$^3$ 4s$^2$ for V and 2s$^2$ 2p$^6$ for O. The k-point sampling of the Brillouin zone was constructed using Monkhorst-Pack scheme [20] with 6×6×6 grids in primitive cell of MgV$_2$O$_6$. The electronic wave function was expanded in a plane-wave basis set with a well converged cutoff energy of 400 eV for all the cases. The equilibrium crystal structures were obtained via geometry optimization in the Broyden-Fletcher-Goldfarb-Shanno (BFGS) minimization method [21]. For obtaining the optimized structure of MgV$_2$O$_6$ the convergence tolerances were set to $1.0 \times 10^{-5}$ eV/atom for the energy, 0.03 eV/Å for the maximum force on atoms, 0.05 GPa for the maximum stress and $1.0 \times 10^{-3}$ Å for the maximum atomic displacement. The elastic stiffness constants of orthorhombic MgV$_2$O$_6$ were obtained by the stress-strain method [22] at the optimized structure under the condition of normal pressure. In this case the criteria of convergence were set to $2.0\times10^{-6}$ eV/atom for energy, $2.0\times10^{-4}$ Å for maximum ionic displacement and 0.006 eV/Å for maximum ionic force. For this investigation the value of the maximum strain amplitude was set to be 0.003. The Debye temperature of MgV$_2$O$_6$ was calculated using our elastic constant data systematically.

## 3. Results and Discussion

### 3.1. Structural Properties

MgV$_2$O$_6$ typically possesses orthorhombic crystal structure with the space group Pbcn (No. 60) and has equilibrium lattice parameters a = 13.610 Å, b = 5.581 Å and c = 4.857 Å [23]. The considered atomic positions of MgV$_2$O$_6$ are given as: Mg (0 0.1596 0.250); V (0.1603 0.3183 0.7716); O(1) ( 0.0788 0.3783 0.4122); O(2) (0.0847 0.1149 0.9198); O(3) ( 0.2539 0.1312 0.6146). The optimized conventional orthorhombic cell and primitive cell of MgV$_2$O$_6$ are shown in Fig.1 (a & b) respectively. The calculated lattice constants and unit cell volume as determined from zero pressure are listed in Table1, along with results of previous experimental data. From Table 1 it is seen that, our calculated lattice parameters are slightly different from the experimental values. This is because our calculated data are simulated at 0 K, while the experimental data are measured at room temperature. These minor deviations from the experimental values evidently bear the reliability of our present DFT based calculations.

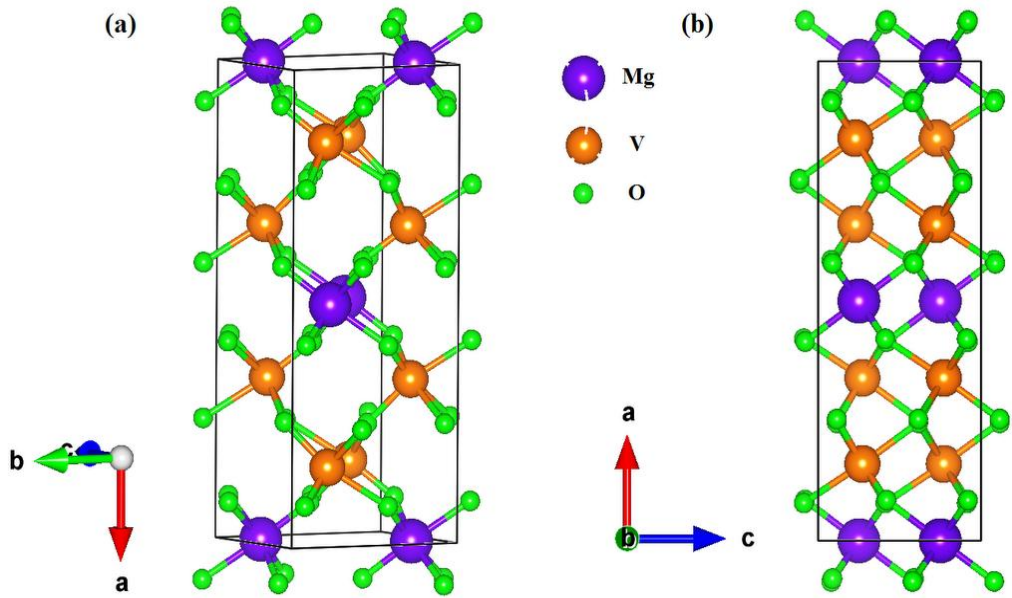

**Fig.1:** The crystal structures of MgV$_2$O$_6$, (a) the three dimensional orthorhombic cell and (b) the two dimensional unit cell.

Table 1: The calculated equilibrium lattice constant "$a$, $b$, $c$", unit cell volume "$V_0$" and bulk modulus "$B_0$" of MgV$_2$O$_6$ semiconductor.

| Properties | Expt.[23] | Other Calculation | Present Calculation | Deviation from Expt. (%) |
|---|---|---|---|---|
| $a, b, c$ (Å) | 13.610, 5.581, 4.857 | - | 13.759, 5.679, 4.856 | 1.09, 1.75, 0.02 |
| $V_0$ (Å$^3$) | 368.93 | - | 379.51 | 2.87 |
| $B_0$ (GPa) | - | - | 160.53 | - |

## 3.2 Elastic Properties

Elastic properties of materials are very significant for the reason that they verify the mechanical stability, ductile or brittle behavior based on the analysis of elastic constants $C_{ij}$, bulk modulus, $B$ and shear modulus, $G$. For example, the bulk modulus $B$ [24], measure the resistance of the volume variation in a solid material and provide an estimation of the elastic response of the materials under the hydrostatic pressure. The shear modulus $G$ [25], illustrate the resistance of a material upon shape change. Also the elastic constant of solids provides a connection between the mechanical and dynamical manner of crystals and confer important information regarding the nature of the forces operating in solids. They also relate the various fundamental solid-state properties such as interatomic potentials, equation of states, specific heat, phonon spectra, thermal expansion, Debye temperature, melting points and Grüneisen parameter [26, 27]. In particular, they provide the information on stability and stiffness of materials. Elastic constants are defined as a Taylor expansion of the total energy, namely the derivative of the energy as a function of a lattice strain [28-34]. The orthorhombic crystal system has nine independent elastic constants: $C_{11}$, $C_{22}$, $C_{33}$, $C_{44}$, $C_{55}$, $C_{66}$, $C_{12}$, $C_{13}$ and $C_{23}$. The required of mechanical stability in an orthorhombic system, which depends on its independent elastic constants leads to the following equations [35, 36],

$C_{11} > 0$; $C_{22} > 0$; $C_{44} > 0$; $C_{33} > 0$; $C_{55} > 0$; $C_{66} > 0$; $C_{11} + C_{22} > 2C_{12}$; $C_{22} + C_{33} > 2C_{23}$;
$C_{11} + C_{33} > 2C_{13}$; $C_{11} + C_{22} C_{33} + 2C_{12} + 2C_{23} + 2C_{13} > 0$.

4The present elastic constants listed in Table 2 are positive and satisfy the stability conditions. This suggests that the orthorhombic $MgV_2O_6$ is mechanically stable compound. To best of our knowledge no reports on the elastic constants of this compound have been found so far, so it is not possible to compare these results with either experimental or theoretical data from other sources.

**Table 1:** The calculated elastic constants of orthorhombic $MgV_2O_6$ (in GPa).

| Material | $C_{11}$ | $C_{12}$ | $C_{13}$ | $C_{22}$ | $C_{23}$ | $C_{33}$ | $C_{44}$ | $C_{55}$ | $C_{66}$ | Ref. |
|---|---|---|---|---|---|---|---|---|---|---|
| $MgV_2O_6$ | 244.84 | 100.25 | 112.21 | 206.62 | 103.70 | 251.24 | 78.74 | 97.59 | 38.59 | Present work |

When single crystal samples are not available a trouble arises because at this moment it is not possible to measure the individual elastic constants. As an alternative, the polycrystalline bulk modulus (B) and shear modulus (G) can be determined. Two approximation methods are available to calculate the polycrystalline modulus, namely, the Voigt method [37] and the Reuss method [38]. For orthorhombic crystal system, the Voigt bulk modulus ($B_V$) and the Reuss bulk modulus ($B_V$) are defined as

$$B_V = \frac{1}{9}(C_{11} + C_{22} + C_{33}) + \frac{2}{9}(C_{12} + C_{13} + C_{23}) \qquad (1)$$

$$B_R = [(S_{11} + S_{22} + S_{33}) + 2(S_{12} + S_{13} + S_{23})]^{-1} \qquad (2)$$

and the Voight shear modulus ($G_V$) and Reuss shear modulus ($G_R$) are

$$G_V = \frac{1}{15}(C_{11} + C_{22} + C_{33} - C_{12} - C_{13} - C_{23}) + \frac{1}{5}(C_{44} + C_{55} + C_{66}) \qquad (3)$$

$$G_R = 15[(S_{11} + S_{22} + S_{33}) - 4(S_{12} + S_{13} + S_{23}) + 3(S_{44} + S_{55} + S_{66})]^{-1} \qquad (4)$$

In Eqs. (2) and (4), $S_{ij}$ is the inverse matrix of the elastic constants matrix $C_{ij}$, which is given by [39]:

$$S_{11} = (C_{22}C_{33} - C_{23}^2)/(C_{11}C_{22}C_{33} + 2C_{12}C_{13}C_{23} - C_{11}C_{23}^2 - C_{22}C_{13}^2 - C_{33}C_{12}^2) \qquad (5)$$

$$S_{12} = (C_{13}C_{23} - C_{12}C_{33})/(C_{11}C_{22}C_{33} + 2C_{12}C_{13}C_{23} - C_{11}C_{23}^2 - C_{22}C_{13}^2 - C_{33}C_{12}^2) \qquad (6)$$

$$S_{13} = (C_{12}C_{23} - C_{22}C_{13})/(C_{11}C_{22}C_{33} + 2C_{12}C_{13}C_{23} - C_{11}C_{23}^2 - C_{22}C_{13}^2 - C_{33}C_{12}^2) \quad (7)$$

$$S_{22} = (C_{11}C_{33} - C_{13}^2)/(C_{11}C_{22}C_{33} + 2C_{12}C_{13}C_{23} - C_{11}C_{23}^2 - C_{22}C_{13}^2 - C_{33}C_{12}^2) \quad (8)$$

$$S_{23} = (C_{12}C_{13} - C_{11}C_{23})/(C_{11}C_{22}C_{33} + 2C_{12}C_{13}C_{23} - C_{11}C_{23}^2 - C_{22}C_{13}^2 - C_{33}C_{12}^2) \quad (9)$$

$$S_{33} = (C_{11}C_{22} - C_{12}^2)/(C_{11}C_{22}C_{33} + 2C_{12}C_{13}C_{23} - C_{11}C_{23}^2 - C_{22}C_{13}^2 - C_{33}C_{12}^2) \quad (10)$$

$$S_{44} = \frac{1}{C_{44}} \quad (11)$$

$$S_{55} = \frac{1}{C_{55}} \quad (12)$$

$$S_{66} = \frac{1}{C_{66}} \quad (13)$$

By taking energy considerations Hill [40] shown that the Voigt and Reuss equations represent the upper and lower limits of true polycrystalline constants, and suggested that a practical estimate of the bulk and shear moduli were the arithmetic means of the extremes. Therefore, for the polycrystalline material the elastic moduli can be approximated by Hill's average and for bulk moduli it is defined as

$$B_H = \frac{1}{2} B_V + B_R \quad (14)$$

and for shear moduli it is defined as

$$G_H = \frac{1}{2} G_V + G_R \quad (15)$$

Again, from the calculated values of bulk modulus $B$ and shear modulus $G$, we can determine the Young's modulus $Y$ and Poisson's ratio v by the following equations [41, 42]:

$$E = \frac{9BG}{G + 3B} \quad (16)$$

$$v = \frac{3B - 2G}{2(3B + G)} \quad (17)$$

Using the relations given above the calculated bulk modulus $B_V$, $B_R$, $B$, shear modulus $G_V$, $G_R$, $G$, Young's modulus $Y$, compressibility $K$, and Poisson's ratio v are listed in Table 2.

According to Pugh's criteria [44, 45], a material should be brittle or ductile if the ratio $B/G$ is less or high than 1.75. From our calculations we see that the value of $B/G$ is higher than 1.75, therefore the material $MgV_2O_6$ behaves in a ductile manner. The information of the Young's modulus and Poisson's ratio are very significant for the industrial and technological applications. Young's modulus is defined by the ratio of stress and strain and used to provide a measure of the stiffness of the solid. The material is stiffer if the value of Young's modulus, $Y$ is high. In this context, due to the higher value of Young's modulus (187.54 GPa) we can be said that the compound under study is relatively stiffer. The value of the Poisson's ratio indicates the degree of directionality of the covalent bonds. For covalent materials the value of is small ($v=0.1$), while for the ionic materials a typical value of $v$ is 0.25[46]. In this case, for $MgV_2O_6$ the calculated Poisson's ratio is about 0.291. Hence, for this compound the ionic contribution to the interatomic bonding is dominated. For the central-force solid materials the lower and upper limits of $v$ are 0.25 and 0.50 respectively [47]. For $MgV_2O_6$, the value of Poisson's ratio is close to 0.291, which indicates that the interatomic force is central force.

Table 2: The calculated Bulk moduli ($B_R$, $B_V$, $B$ in GPa), shear moduli ($G_R$, $G_V$, $G$ in GPa), Young's modulus ($Y$ in GPa), Compressibility ($K$ in GPa$^{-1}$), G/B, and Poisson's ratio (v) of orthorhombic $MgV_2O_6$ compared with other theoretical results of similar type of compound.

| Material | $B_V$ | $B_R$ | $B$ | $G_V$ | $G_R$ | $G$ | $Y$ | $K$ | G/B | v | $A_1$, $A_2$, $A_3$ | Ref. |
|---|---|---|---|---|---|---|---|---|---|---|---|---|
| $MgV_2O_6$ | 148.33 | 148.43 | 148.38 | 68.75 | 76.56 | 72.66 | 187.54 | 0.0067 | 0.489 | 0.291 | $A_1 = 2.537$ $A_2 = 1.558$ $A_3 = 0.615$ | Present work |
| $MgBi_2O_6$ | 50.78 | 51.32 | 51.05 | 37.37 | 14.10 | 25.73 | 66.09 | 0.0070 | 0.500 | 0.280 | - | [43] |

Several low symmetry crystals reveal a high degree of elastic anisotropy [48]. The shear anisotropic factors on dissimilar crystallographic planes give a measure of the degree of anisotropy in atomic bonding in different planes. The shear anisotropic factors are defined by the equations:

$$A_1 = \frac{4C_{44}}{C_{11}+C_{33}-2C_{13}} \quad \text{for the \{100\} plane} \tag{18}$$

$$A_2 = \frac{4C_{55}}{C_{22}+C_{33}-2C_{23}} \quad \text{for the \{010\} plane} \tag{19}$$

$$A_3 = \frac{4C_{66}}{C_{11}+C_{22}-2C_{12}} \quad \text{for the \{001\} plane} \tag{20}$$

The calculated values of elastic anisotropy are given in Table 2. A value of unity indicates that the crystal exhibits isotropic properties whereas the values of other than unity represent varying degrees of anisotropy. From Table 2, it can be seen that $MgV_2O_6$ exhibits larger anisotropy in the {100} and {010} planes and this compound exhibits almost isotropic properties for the {001} plane according to other planes.

### 3.3 Electronic properties

The calculations of the electronic properties of materials provide significant information for understanding the physical properties and bonding nature of crustal. Therefore the details study of bonding characteristics of $MgV_2O_6$ is so important. In this case the study of the electronic band structure and the density of states (DOS) play the most significant task. In order to get a deep insight about the bonding nature of this compound, the total density of states (TDOS) and the partial density of states (PDOS) have been calculated and shown in Fig.3.

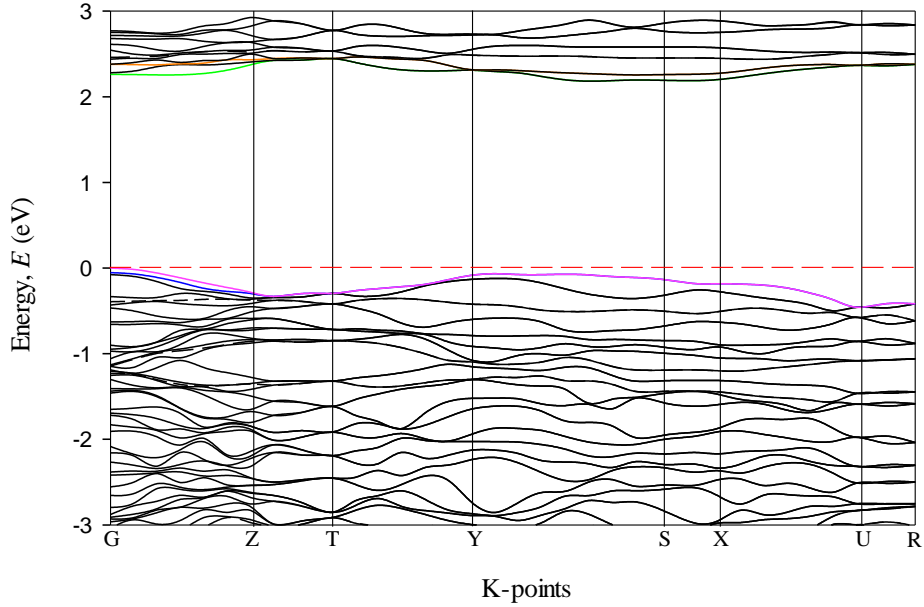

**Fig. 2:** Electronic band structure of $MgV_2O_6$ semiconductor along high symmetry direction in the Brillouin zones.

To identify the metallic nature of $MgV_2O_6$ band structure along the high symmetry direction in the Brillouin zones has been investigated and is presented in Fig.2. Analyzing the band structure diagram as shown in Fig.2, we determine the indirect band gap of $MgV_2O_6$ is 2.195 eV, which reveals that the compound under study is a semiconductor. Fig.3 shows the total and partial density of states of $MgV_2O_6$. The valance band lies between -65.0 eV to 0 eV (Fermi level). In valance band most of the contribution comes from Mg-2p, V-3p and O-2p orbitals. However we observe predominant hybridization for V-3d and O-2s states near the Fermi level. In conduction band the similar kind of domination of V-3d and O-2s states is noticed. The coincidence between V-3d and O-2s states suggests that there exist covalent bond between those orbital. However at Fermi level O-2s orbital contributes the most, which are the common characteristics of oxide semiconducting materials.

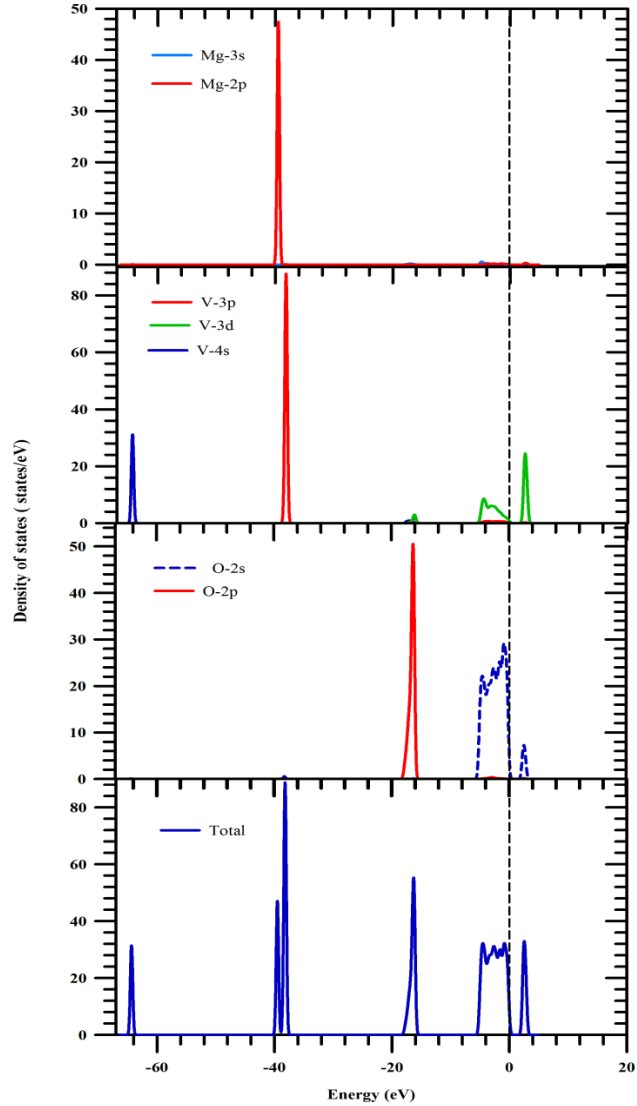

**Fig. 3:** The partial and total density of states of MgV$_2$O$_6$.

### 3.4 Optical Properties

The detailed study of the optical properties is very essential because it helps us to get better understanding of the material's electronic structure. The optical properties of a solid can be efficiently explained in a unique manner by its complex dielectric function $\varepsilon$, which represents the linear response of the system to an external electromagnetic field with a small wave vector. It can be expressed by the equation $\varepsilon(\omega) = \varepsilon_1(\omega) + i\varepsilon_2(\omega)$, which is closely related to the electronic structures. After calculating the electronic structure, the optical properties can be calculated in CASTEP in a straightforward way. The imaginary part $\varepsilon_2(\omega)$ of a dielectric function $\varepsilon(\omega)$ can be determined by direct numerical evaluation of matrix elements of the electric dipole operator between the occupied states in the valence band (VB) and empty states in the conduction band (CB) [49]:

$$\varepsilon_2(\omega) = \frac{2e^2\pi}{\Omega\varepsilon_0} \sum_{k,v,c} \left|\langle \psi_k^c | \hat{u} \cdot \vec{r} | \psi_k^v \rangle\right|^2 \delta(E_k^c - E_k^v - E) \qquad (21)$$

where $\omega$ is the frequency of light, $e$ is the charge of electron, $\hat{u}$ is the vector defining the polarization of the incident electric field, and $\psi_k^c$ and $\psi_k^v$ are the conduction and valence band wave functions at $k$, respectively, $E = \hbar\omega$ is the incident energy of photon, and $\varepsilon_0$ is the dielectric permittivity of vacuum.

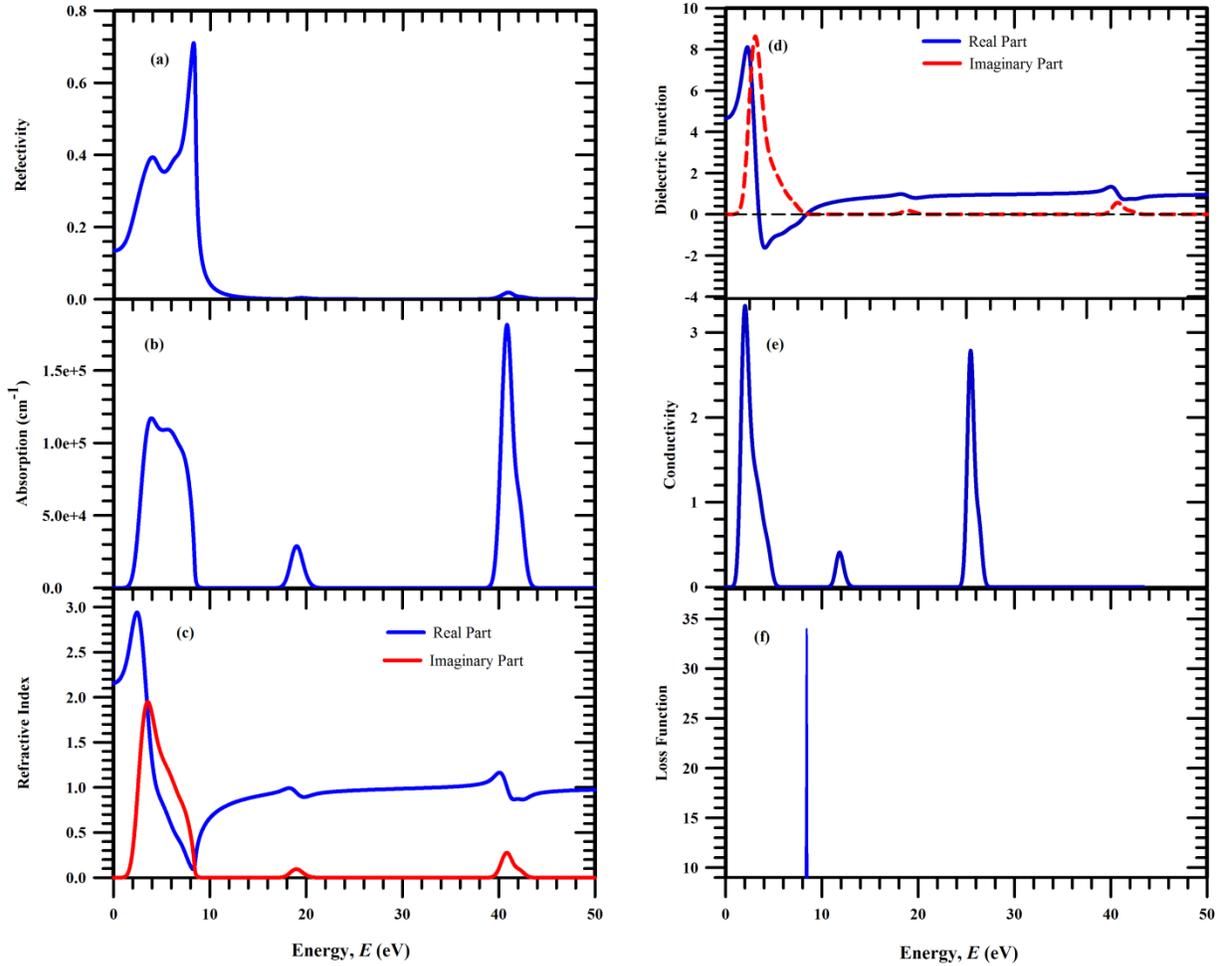

**Fig.4**: The optical functions (a) reflectivity, (b) absorption, (c) refractive index, (d) dielectric function, (e) conductivity and (f) loss function of $MgV_2O_6$ for polarization vector [100].

From the imaginary part $\varepsilon_2(\omega)$ the real part $\varepsilon_1(\omega)$ of the complex dielectric function is obtained using the Kramers-Kronig relations [50-52]:

$$\varepsilon_1 = 1 + \frac{2}{\pi} M \int_0^\infty \frac{\omega' \varepsilon_2(\omega') d\omega'}{(\omega'^2 - \omega^2)} \qquad (22)$$

Where, M indicates the principle value of the integral. With the knowledge of the complex dielectric function, all other frequency dependent optical functions, such as refractive index, absorption spectrum, loss-function, reflectivity and conductivity (real part) are derived from the formalism developed in Ref. [49].

The optical functions of compound $MgV_2O_6$ are calculated for photon energies up to 50 eV to the direction of polarization vector [100]. We have used a 0.5 eV Gaussain smearing for all calculations for the reason that this smears out the Fermi level in such a manner that *k*-points will be more efficient on the Fermi surface. Fig. 4(a) shows the reflectivity spectrum of $MgV_2O_6$ as a function of photon energy. The reflectivity starts with a value of 0.144 and then rises to reach maximum value of 0.717 between 4.18 and 8.56 eV. The reflectivity value drops in the high energy region as a result of intraband transition. The large reflectivity for E<1eV indicates the characteristics of high conductance in the low energy region. The analysis shows that the compound $MgV_2O_6$ would be used as a good coating material.

The absorption coefficient offers very essential information about the optimum solar energy conversion efficiency. It indicates how far light of a specific energy or wavelength can penetrate into the material before being absorbed. The absorption spectrum of $MgV_2O_6$ in direction [100] is shown in Fig. 4 (b). It observed that the spectra start from energy 1.46 eV which indicates the semiconducting nature of this compound. This has also been confirmed from the band structure of $MgV_2O_6$ (Fig.2). In the absorption spectra several peaks are observed but the highest peak located at 40.75 eV indicating good absorption coefficients in the ultraviolet region for $MgV_2O_6$.

The refractive index is very essential optical function of a material which determines how much light is bent or refracted when entering a material. Fig. 4 (c) exhibits the refractive index of MgV$_2$O$_6$ as a function of photon energy. The static refractive index of MgV$_2$O$_6$ is 2.95. Hence it is evident from figure that refractive index is high in the infrared region and gradually decreased in the visible and ultraviolet regions.

Dielectric function is the most general properties of solids and it describes the absorption and polarization properties of materials. As a function of photon energy the calculated real and imaginary parts of dielectric function of MgV$_2$O$_6$ are shown in Fig. 4 (d). It is observed that the value of $\varepsilon_2(\omega)$ becomes zero at about 62 eV indicating that this materials become transparent above this certain value. When the value of $\varepsilon_2(\omega)$ becomes nonzero, then absorption begins. Therefore it is clear from Fig. 4 (d) that absorption occurs among 1.52-8.4 eV, 17.36-20.27 eV and 39.02-43.90 eV, which is also evident from Fig. 4 (b).

Fig.4 (e) represents the conductivity spectra of MgV$_2$O$_6$ as a function of photon energy. It is evident from the figure that the photoconductivity starts with non-zero photon energy which indicates that the compound under study has band gap and exhibits semiconducting nature. This is also evident from the band structure calculation (Fig. 2). It is seen that the value of conductivity is high in the low energy region and decreases in the high energy region. There is no photoconductivity when the photon energy is above 43 eV.

The energy loss spectrum is a crucial optical parameter to explain the energy loss of a fast electron when traversing the material and is large at the plasma frequency [53]. Fig. 4 (f) shows the loss function of MgV$_2$O$_6$ as a function of photon energy. It is observed that the highest peak is found at 8.42, which indicates rapid reduction in the reflectance. This prominent peak is called bulk plasma frequency of material, which appears at $\varepsilon_2$ <1 and $\varepsilon_1$ = 0 respectively [54-56]. Therefore from energy loss spectrum, we observed that the plasma frequency is equal to 8.42 eV. Hence this material becomes transparent when the incident photon energy is higher than 8.42 eV.

### 3.5. Debye temperature

The Debye temperature is particularly important thermodynamic quantity for the measurement of various physical properties of crystal such as thermal expansion, specific heat, melting point etc. The highest temperature that can be attained due to a single normal lattice vibration is generally referred to as the Debye temperature and is denoted as $\Theta_D$. There are different methods available for determining the value of $\Theta_D$. In this article we have used the calculated elastic constants data to evaluate the Debye temperature of MgV$_2$O$_6$. In this way the value of $\Theta_D$ can be obtained directly by using the following equation [57],

$$\theta_D = \frac{h}{k_B}\left(\frac{3N}{4\pi V}\right)^{\frac{1}{3}} \times v_m \tag{23}$$

Where, h is the Planck constant, N is the number of atoms in unit cell and V is the volume of the unit cell. The detailed procedures of determining the value of $V_m$ is discussed elsewhere from Eq. 16-18 [58]. The evaluated Debye temperature of MgV$_2$O$_6$ using equation (23) is 656.99 K where $V_m$ is 4833.52 m/s.

## 4. Conclusion

In summary, in this present research work we have investigated the structural, elastic, electronic, optical and thermodynamic properties of orthorhombic MgV$_2$O$_6$ via *ab-initio* evolutionary simulation method. We have found that the lattice parameters obtained after optimization are in agreement with the experimental findings. We have calculated the independent elastic constants and derived the bulk, shear and Young's modulus, Poisson coefficients, compressibility and Poisson's ratio. The results show that the orthorhombic MgV$_2$O$_6$ is mechanically stable, anisotropic and behave in a brittle manner. According to the value of Poisson's ratio we have observed that this compound has some ionic feature. The electronic band structure shows that orthorhombic MgV$_2$O$_6$ has indirect band gap with 2.195 eV. Further the first time study of optical properties such as reflectivity, absorption coefficient, refractive index, dielectric function, photoconductivity and energy-loss function are determined and analyzed in detail. The study of photoconductivity reveals the semiconducting nature of MgV$_2$O$_6$. The larger reflectivity in the low energy region indicates suitability of the compound for use in the solar cell as a coating material to remove solar heating. We have also determined the Debye temperature using the calculated elastic constants data and the evaluated Debye temperature of MgV$_2$O$_6$ is 656.99 K.